# Plasmonic Ferroelectric Modulators

Andreas Messner , Felix Eltes , Ping Ma , Stefan Abel , Benedikt Baeuerle , Arne Josten , Wolfgang Heni , Daniele Caimi, Jean Fompeyrine , and Juerg Leuthold , *Fellow, IEEE, Fellow, OSA*

*(Invited Paper)*

*Abstract*—Integrated ferroelectric plasmonic modulators featuring large bandwidths, broad optical operation range, resilience to high temperature and ultracompact footprint are introduced. Measurements show a modulation bandwidth of 70 GHz and a temperature stability up to 250 °C. Mach–Zehnder interferometer modulators with 10-$\mu$m-long phase shifters were operated at 116 Gbit/s PAM-4 and 72 Gbit/s NRZ. Wide and open eye diagrams with extinction ratios beyond 15 dB were found. The fast and robust devices are apt to an employment in industrial environments.

*Index Terms*—Electrooptic modulators, ferroelectric devices, high-speed integrated circuits, plasmonics.

## I. INTRODUCTION

HIGH-SPEED, compact and power-efficient electro-optic (EO) modulators are currently in the spotlight of research as they are key components in high-capacity optical links. Many physical effects have already been exploited to perform electro-optic (EO) modulation. Among them are for example, the quantum-confined Stark effect [1], [2], the plasma dispersion effect [3], [4] or the linear EO effect, commonly called Pockels effect. Here, the Pockels effect is of particular interest as it provides a large optical bandwidth and a pure phase modulation such as needed to operate phase shifters in MZ and IQ modulators. Both MZ and IQ-modulators are key elements for encoding advanced modulation formats in high-capacity communication systems.

State-of-the-art EO Pockels modulators commonly rely on the Pockels effect of lithium niobate (LiNbO₃, LNB), a ferroelectric material. Transmission rates of 1.6 Tbit/s have already been demonstrated using a LNB IQ modulator [5]. Yet, those LNB modulators are typically based on weakly guiding waveguides, resulting in centimeter-long devices [6]. Also, the long

Manuscript received August 4, 2018; revised October 15, 2018; accepted November 7, 2018. Date of publication November 14, 2018; date of current version February 20, 2019. This work was supported in part by Swiss National Foundation under Project 200021_159565 PADOMO and Project IZCJZ0-158197/1 FF-Photon, in part by the European Commission under Grants FP7-ICT-2013-11-619456-SITOGA, H2020-ICT-2015-25-688579 PHRESCO and H2020-ICT-2017-1-780997 plaCMOS, and in part the Swiss State Secretariat for Education, Research and Innovation under contracts 15.0285 and 16.0001. *(Corresponding author: Ping Ma.)*

A. Messner, P. Ma, B. Baeuerle, A. Josten, W. Heni, and J. Leuthold are with the Institute of Electromagnetic Fields, ETH Zürich, Zürich 8092, Switzerland (e-mail: amessner@ethz.ch; mapi@ethz.ch; bbaeuerle@ethz.ch; ajosten@ethz.ch; wheni@ethz.ch; leuthold@ethz.ch).

F. Eltes, S. Abel, D. Caimi, and J. Fompeyrine are with IBM Research–Zurich, Rüschlikon 8803, Switzerland (e-mail: fee@zurich.ibm.com; sab@zurich.ibm.com; cai@zurich.ibm.com; jfo@zurich.ibm.com).

Color versions of one or more of the figures in this paper are available online at http://ieeexplore.ieee.org.

Digital Object Identifier 10.1109/JLT.2018.2881332

travelling wave electrodes suffer from high electrical losses at high frequencies and the walk-off between electrical and optical signals [6], both of which ultimately limit the modulation bandwidth and power efficiency of the devices.

The need of higher integration densities led to the emerging of the silicon (Si) photonics platform. Si photonics was proposed in 1985 [3]. It relies on silicon-on-insulator (SOI) wafers and promises to leverage scaling effects similar to the mature CMOS technology and more compact footprint due to large contrasts in refractive indices between adjacent layers. However, Si does not exhibit the Pockels effect so that Si modulators commonly resort to the plasma dispersion effect. This effect faces a tradeoff between modulation speed and strength and acts on both phase and amplitude of the optical signal. Therefore, a functional material which exhibits the Pockels effect and which can be cointegrated with the Si photonics platform is of great interest for the development of high performance EO modulators.

A possible research direction towards this aim is the silicon-organic-hybrid (SOH) platform [7], [8]. There, functional organic materials offering the Pockels effect are introduced and modulators have already demonstrated bandwidths of 100 GHz [9] and data rates of up to 400 Gbit/s [10].

Another technology is the "lithium niobate on insulator" (LNOI) platform, which is based on a thin film of LNB on a thick insulating SiO₂ layer. Being proposed in 2010 [11], it has attracted an increasing attention [12]–[14]. Due to the sufficient refractive index contrast between LNB ($n_{LNB} \approx 2.2$ at $\lambda = 1550$ nm) and SiO₂ ($n_{SiO_2} \approx 1.44$), this platform omits Si completely and still allows miniaturization of basic passive photonic components, such as ridge waveguides, Y-splitters, multimode interference couplers (MMIs), and resonators [12], [15], [16]. Active components such as EO modulators are directly enabled by the Pockels effect of LNB and benefit from the platform's strong modal confinement and effective overlap between optical and electric signals [12]. Additionally, the low-permittivity SiO₂ layer facilitates velocity matching between the two signals to further increase the modulation bandwidth. Recently, Zhang *et al.* reported a 100 GHz bandwidth Mach-Zehnder modulator (MZM) with 5 mm long phase shifters and a voltage length product of $V_\pi L = 2.2$ Vcm [14]. While the LNB platform has made significant progress beyond state-of-the-art, one of the main limitations of the technology is the small substrate size (<6 inch), which is incompatible with current CMOS standards. In addition, advanced device designs featuring smaller footprint, co-integration with Si electronics and even lower power consumption deserve more research efforts.

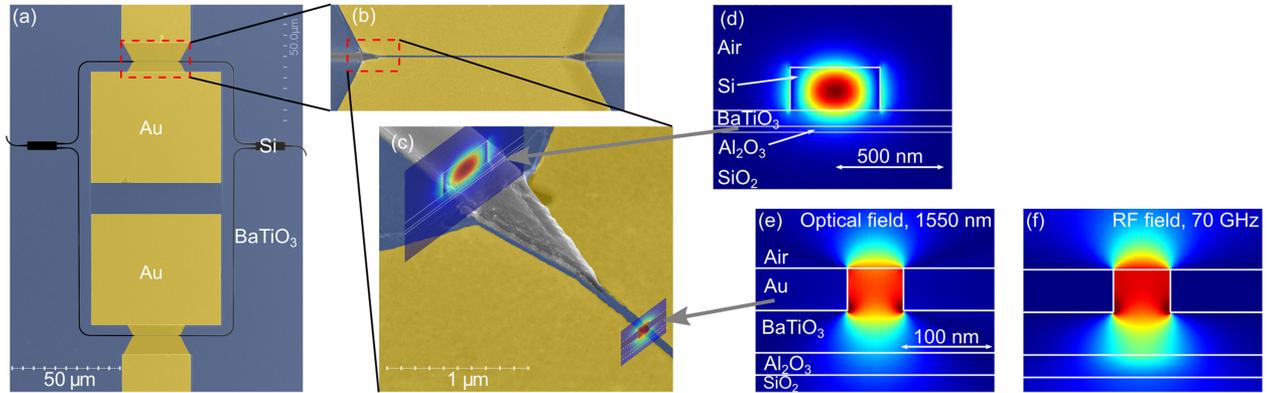

Fig. 1.    False-color SEM images of (a) Plasmonic ferroelectric Mach-Zehnder modulator; (b) a plasmonic phase modulator. (c) Close-up of a tapered mode converter. The Si bus waveguide and tapered mode converter are used to couple the photonic mode into the Au-BTO-Au plasmonic slot waveguide. Insets: Simulated photonic and plasmonic mode profiles, respectively. (d) Simulated electrical field of the photonic mode in the Si waveguide (TE mode). (e) Simulated electrical field of the plasmonic mode (transverse component). (f) Simulated RF field in the slot waveguide (transverse component). We assume $\epsilon_{r,\mathrm{BTO}} = 1000$.

And indeed, there are two opportunities to further reduce the voltage-length product and simultaneously increase the EO bandwidth: The first is to choose a material with a higher Pockels coefficient than LNB and the second is to switch to a plasmonic device geometry which could provide higher modulation efficiency and smaller $RC$ time constant than typical photonic devices.

The first opportunity to decrease the voltage length product arises from switching from LNB with a Pockels coefficient of only 30 pm/V [17] to other ferroelectric materials that exhibit a stronger Pockels effect. Barium titanate (BaTiO$_3$, BTO) is among the most promising ferroelectric materials for EO applications, because of its strong Pockels effect [18]. Its largest Pockels tensor element is reported to be $r_{42} = 1300$ pm/V in the unclamped, zero-stress case, and to be $r_{42} = 700$ pm/V in the clamped case [19]. The Pockels effect of BTO has already been exploited in active Si photonic devices [20]–[24] and [53], since BTO can be epitaxially grown on Si [20]. Recently, EO modulators based on BTO have even been monolithically integrated on an advanced Si photonic platform, opening up possibilities towards monolithic integration with electric circuits in a foundry environment [24].

The second opportunity to decrease the voltage-length product and simultaneously raise the EO bandwidth dramatically is to transit from photonics to plasmonics. A plasmonic modulator device can be built using a metal-insulator-metal slot waveguide geometry. Here, metal electrodes are used both as guiding structures for the optical mode and as radio frequency (RF) electrodes for the electrical signal. The slots can be filled with nonlinear functional materials introducing the Pockels effect. This leads to the so-called plasmonic-organic hybrid (POH) approach, which adopts nonlinear organic materials filled into a narrow, plasmonic metallic slot waveguide. The POH technology offers enhanced modulation efficiency and bandwidth [25], [26]: To this point successful high-speed modulation has been demonstrated on a few µm²-footprints [27]–[29] with high-speed data modulation of up to 200 Gbit/s [30] that can be related to a flat EO frequency response up to 325 GHz [31], [32] and a record low power consumption of as little as 2.8 fJ/bit at 100 Gbit/s

[33]. While these demonstrations are very successful, organic materials encounter reservations by the industry as a CMOS compatible solution of thermally stable materials for low-cost mass-produced transmitter products.

In this paper, we report on the current progress in realizing ultra-compact and high-speed plasmonic ferroelectric modulators (PFMs) by effectively combining the epitaxially grown functional ferroelectric BTO material with a plasmonic device design. We present 72 Gbit/s non-return-to-zero (NRZ) data modulation and 116 Gbit/s 4-level pulse-amplitude modulation (PAM-4) experiments. Experiments are performed with modulators featuring an active section that is as short as 10 µm, a measured frequency response that exceeds 70 GHz (only limited by the experimental instruments) and a record low voltage-length product between 150 and 200 Vµm, depending on the modulation frequency. Furthermore, temperature stability up to 250 °C is shown.

The paper is organized as follows. Section II describes the device design and concept. Section III depicts the fabrication process of the device. Section IV discusses the influence of BTO's crystalline orientation on the device design. A derivation of the effective Pockels coefficients is given. Section V presents passive as well as EO characterization experiments. These results include the frequency response of PFMs and the temperature stability test. Section VI shows the data modulation experiments of 116 Gbit/s PAM-4 and 72 Gbit/s NRZ. Section VII concludes the work by summarizing the key results.

## II. DEVICE CONCEPT AND DESIGN

This section presents the concept and design of the efficient, compact, and ultra-fast modulators. Two plasmonic ferroelectric phase shifters are integrated into a photonic Mach-Zehnder interferometer (MZI). Photonic elements are employed for fiber-to-chip coupling and on-chip routing.

### A. Plasmonic Ferroelectric Mach-Zehnder Modulators

Fig. 1(a) shows a fabricated PFM, which consists of two ferroelectric phase shifters, Fig. 1(b), in a MZI. We employ Si photonic grating couplers [34] to couple the light to waveguides on the chip and multimode interference couplers (MMIs) to split up the light into the two arms of the PFM. A tapered mode converter, Fig. 1(c), couples light from the 450-nm-wide Si access waveguide, Fig. 1(d), to the plasmonic mode, Fig. 1(e), and at the end of the phase shifter the plasmonic mode is coupled back to the Si waveguide by another mode converter. Both the photonic and the plasmonic mode are TE-polarized. The asymmetry in the MZI arm lengths (100 $\mu$m difference) allows for an easily accessible evaluation of the fabricated devices and supplements the capability of the voltage-controllable operating point by offering an additional spectral tunability.

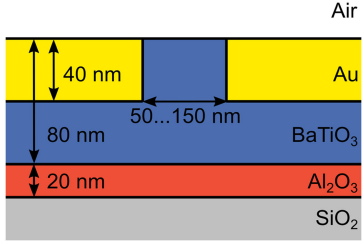

Fig. 2. A cross section through the plasmonic ferroelectric phase shifter. 80 nm BTO is wafer-bonded to 3 $\mu$m SiO$_2$ on Si, using a 20-nm-thick Al$_2$O$_3$ buffer layer. Two gold electrodes enclose a narrow $(50 \ldots 150$ nm) and 40-nm-tall BTO slab.

Plasmonic ferroelectric phase modulators are the key components of our devices. They are based on an Au-BTO-Au plasmonic slot waveguides, a concept which has already been successfully demonstrated with organic EO materials [25], [35]. A schematic cross-section of such a phase modulator is shown in Fig. 2. The metal electrodes form a very narrow $(50 \ldots 150$ nm) BTO-filled slot, and act both as optical waveguides and as feeds for the RF field.

### B. Characteristics of Plasmonic Ferroelectric Modulators

PFMs feature efficient modulation, low drive voltages, a most compact footprint while offering broadband frequency response.

The PFMs are efficient because they benefit from three distinct advantages over more conventional modulators. First, they benefit from BTO as an EO material, which provides one of the highest Pockels coefficients among all known materials [18]. Second, they benefit from an almost perfect overlap between the EO material, the optical mode ($n_{BTO} \approx 2.27$ at $\lambda = 1550$ nm [36]) and the electrical field, ($\varepsilon_{r,BTO} \approx 1000$, measured at 20 GHz [37], [38]), as shown in Fig. 1(e) and (f), respectively. Third, a high group index leads to field enhancement due to a slow-down effect. All of these effects strongly change the effective refractive index $n_{eff}$ of the plasmonic mode,

$$\Delta n_{eff} = \Delta n_{BTO} \cdot \Gamma \tag{1}$$

Here, $\Delta n_{BTO}$ is BTO's refractive index change, which is proportional to its Pockels coefficient, and $\Gamma$ is the field interaction factor [7], [8]. The $\Gamma$-factor does not only comprise the strong field confinement and overlap, but additionally reflects the slow-down effect, because it is inversely proportional to the group velocity $v_g$, $\Gamma \propto n_g^{-1}$ [7]. In plasmonic structures such as discussed here, the field interaction factor is typically larger than 1, which means that the material response is actually amplified by the waveguiding structure.

Low drive voltages – accompanied by low power consumption – can as well be traced back to a few geometrical advantages of the concept. For instance the small wavelength of plasmons allows for the guiding in narrow slot waveguides. This then provides another advantage. Thanks to these narrow slots (width: $d$), one obtains a very strong electrical field $E = V/d$, which is the key stimulus for the Pockels effect, described by the Pockels coefficient $r$.

$$\Delta n_{BTO} \propto r \cdot E = r \cdot \frac{V}{d} \propto \frac{1}{d} \tag{2}$$

Voltage requirements to operate a PFM are further relaxed by the PFM's open-circuit behavior and the push-pull configuration of the MZM. Because their parasitic impedance can be neglected with respect to 50 $\Omega$ driving electronics, the devices appear as open-circuit terminated, and reflect the incoming RF signal. This leads to a 6 dB voltage gain as compared to a terminated device with travelling wave electrodes. Drivers that withstand the reflected wave are commercially available and have been successfully used in our experiment. Another factor two enhancement in voltage is gained from joining two plasmonic phase shifters to a MZM. The nature of the linear EO effect allows for operating the MZM in a push-pull configuration: Phase modulations of opposite signs are generated in the two arms and added up at the output. While this halves the needed driving voltage, it also results in a pure, inherently chirp-free modulation.

Thanks to this high efficiency, a single phase shifter is as short as 10 $\mu$m and does not require travelling wave electrodes. Its footprint can be as small as 20 $\mu$m$^2$, although it is currently constrained by the size of the electric contact pads.

Plasmonic modulators offer ultra-broad EO bandwidths: The compactness dramatically reduces the devices' parasitic capacitance $C$. Due to the metallic RF feeds, the devices also feature a very small parasitic resistance $R$. As a result, plasmonic slot modulators offer a uniquely small $RC$ time constant, which enables EO bandwidths pushing to the THz regime [26], [39]. In strong contrast to carrier effects, the Pockels effect is an ultra-fast field effect and does not jeopardize the device's frequency response [40].

## III. FABRICATION PROCESS

The devices discussed here are fabricated on a Si/3 $\mu$m SiO$_2$/20 nm Al$_2$O$_3$/80 nm BTO/220 nm Si wafer. To manufacture this layer stack, an 80-nm-thick BTO film was deposited epitaxially on a SrTiO$_3$-coated SOI wafer by molecular beam epitaxy (MBE). The 4-nm-thin SrTiO$_3$ layer is necessary to enable an epitaxial relationship between Si (100) and BTO [20]. Using 10-nm-thick Al$_2$O$_3$ adhesion layers, the wafer was bonded [41] to a Si wafer with 3-$\mu$m-thick thermal SiO$_2$ cladding. The original BTO handle wafer and its buried oxide (BOX) layer were stripped. The resulting layer stack resembles a standard SOI wafer with a functionalized BOX layer directly adjacent to the device Si layer.

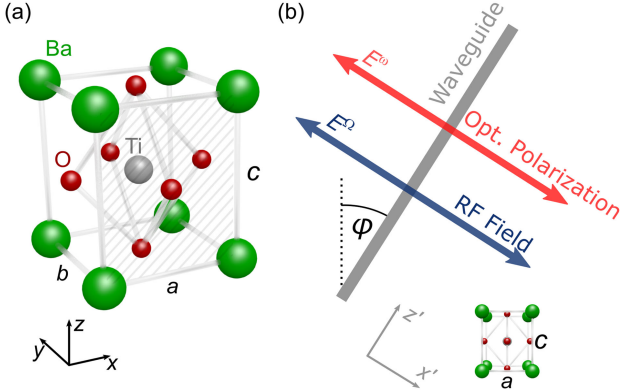

Fig. 3. (a) Crystal structure of the BTO unit cell in the tetragonal phase. The $a$- and $b$-axis have the same length, the $c$-axis is slightly longer. The titanium atom resides in a double-well potential whose minima are slightly shifted from the volume center along the $c$-axis. This gives rise to BTO's spontaneous polarization, ferroelectricity and its Pockels effect. (b) Orientation of waveguide within BTO crystal as discussed in this work. The waveguide lies in the $(ac)$-plane. Its direction is offset by an angle $\varphi$ with respect to the $c$-axis. The RF field $E^\Omega$ and the optical polarization are oriented transversely to the waveguide direction.

E-beam lithography (EBL) was used for all resist exposures. The Si photonic components were structured by a chemical dry etching process. Subsequently, 50-nm-wide and 40-nm-tall BTO strips were structured by a physical dry etching process. The Au electrodes were applied by e-beam evaporation and a lift-off process [23].

## IV. Effects of Crystal Structure on Pockels Effect

The crystalline structure of BTO determines the nonlinear EO response of the material. In this section, we first discuss the EO properties of the bulk single crystal and calculate the effective Pockels coefficient related to an optical wave. Second, we calculate the effective Pockels coefficient of the MBE-grown BTO thin-film considering its multi-domain nature and compare the extracted coefficients to the values reported in the literature.

### A. The BTO Bulk Crystal

Fig. 3(a) shows the unit cell of BTO. At room temperature, BTO crystallizes in a tetragonal, pseudo-Perovskite structure, where the $a$- and $b$-axis of the crystal have the same length, while the $c$-axis is slightly longer. In bulk, the lattice constants are $a = b = 3.991$ Å and $c = 4.035$ Å [42]. The titanium atom in the volume center of the unit cell is displaced by 0.02 Å along the $c$-axis [42], which gives rise to the spontaneous polarization, the ferroelectricity and the Pockels effect of BTO. Due to its tetragonal lattice, BTO is birefringent, its ordinary refractive index for light polarized along the $a$- or $b$-axis is $n_o = 2.30$, its extraordinary index for light polarized along the $c$-axis is $n_e = 2.27$ [36].

BTO's refractive index change due to its Pockels effect in response to an externally applied electrical field can be described

#### TABLE I
#### BTO Pockels Tensor Components

| Tensor element | Stress-free | Stressed |
|---|---|---|
| $r_{13} = r_{23}$ | $10.2 \pm 0.6$ pm/V | $8 \pm 2$ pm/V |
| $r_{33}$ | $105 \pm 10$ pm/V | $40.6 \pm 2.5$ pm/V |
| $r_{42} = r_{51}$ | $1300 \pm 100$ pm/V | $730 \pm 100$ pm/V |

by using the tensorial expression as follows

$$
\begin{pmatrix}
\Delta(1/n^2)_1 \\
\Delta(1/n^2)_2 \\
\Delta(1/n^2)_3 \\
\Delta(1/n^2)_4 \\
\Delta(1/n^2)_5 \\
\Delta(1/n^2)_6
\end{pmatrix}
=
\begin{pmatrix}
0 & 0 & r_{13} \\
0 & 0 & r_{23} \\
0 & 0 & r_{33} \\
0 & r_{42} & 0 \\
r_{51} & 0 & 0 \\
0 & 0 & 0
\end{pmatrix}
\cdot
\begin{pmatrix}
E_x^\Omega \\
E_y^\Omega \\
E_z^\Omega
\end{pmatrix}
\tag{3}
$$

$E^\Omega$ is the applied external field, $r_{mn}$ are the Pockels coefficients in the Voigt notation [43] ($x = 1, y = 2, z = 3; xx = 1, yy = 2, zz = 3, yz = 4, xz = 5, xy = 6$). The crystallographic point group 4 mm of BTO explains the zero entries of that matrix. Moreover, it is $r_{13} = r_{23}$ and $r_{42} = r_{51}$. Values of the Pockels tensor components reported by Zgonik et al. are given in Table I [19].

The $r_{42}$ and $r_{51}$ components are by far the strongest tensor elements.

The question arises if the EO response can be maximized by changing the angle $\varphi$ between the waveguide and the $c$-axis, see Fig. 3(b). To this end, we calculate the effective Pockels coefficient $r_{\text{eff}}$ in dependence on $\varphi$ and follow the derivation given in ref. [44].

We start by stating the index ellipsoid of BTO, which is equivalent to the tensorial notation of Eq. (3). Furthermore, we use that $r_{13} = r_{23}$ and $r_{42} = r_{51}$:

$$
\left(\frac{1}{n_o^2} + r_{13}E_z\right)x^2 + \left(\frac{1}{n_o^2} + r_{13}E_z\right)y^2 + \left(\frac{1}{n_e^2} + r_{33}E_z\right)z^2
$$
$$
+ (r_{42}E_y)\,2yz + (r_{42}E_x)\,2zx = 1. \tag{4}
$$

The $x$-, $y$- and $z$-axes correspond to crystal's $a$-, $b$- and $c$-axes, see Fig. 3(a). We assume the waveguide to be in the $xz$-plane, as indicated in Fig. 3(b). The coordinates $x'$ and $z'$ give the waveguide's transversal and longitudinal direction, respectively. The coordinate transformation between $(x, z)$ and $(x', z')$ is given by

$$
\begin{pmatrix} x \\ z \end{pmatrix} = \begin{pmatrix} \cos\varphi & \sin\varphi \\ -\sin\varphi & \cos\varphi \end{pmatrix} \begin{pmatrix} x' \\ z' \end{pmatrix}. \tag{5}
$$

The modulating field is applied transversely to the waveguide in $x'$ direction. Hence, the electrical fields $E_x$ and $E_z$ along the crystalline axes can be then expressed by

$$
E_x = E_{x'} \cdot \cos(\varphi)
$$
$$
E_z = -E_{x'} \cdot \sin(\varphi) \tag{6}
$$

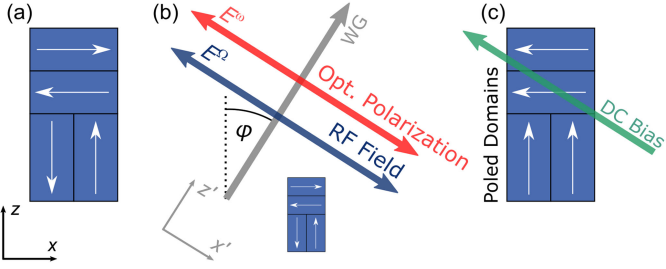

Fig. 4. (a) Top view of a BTO *a*-axis film as grown by MBE on Si. The white arrows indicate the direction of the domains' spontaneous polarization aligned along the *c*-axis. Due to the epitaxial relation to the Si substrate, the domains stand perpendicularly on each other, the direction of their spontaneous polarization is random. (b) A waveguide's orientation in the crystalline thin film, aligned at an angle $\varphi$ with respect to the crystalline domains. The directions of the modulating RF field and the optical polarization are also indicated. (c) A DC bias field aligns the directions of spontaneous polarization of all domains and hence poles the BTO thin film. In this configuration, the domains' electro-optic coefficients do not counteract each other but all contribute to the effective Pockels coefficient.

It is $E_y = E'_y = 0$.

We insert these transformations into the indicatrix, and only consider the term with the $x'^2$ component, which determines how a modulating field in $x'$ direction acts on the refractive index of light polarized in the same direction. We find the term

$$\left[\left(\frac{1}{n^2}\right)_{x'x'} + \Delta\left(\frac{1}{n^2}\right)_{x'x'}\right] = \frac{\cos^2\varphi}{n_o^2} + \frac{\sin^2\varphi}{n_e^2}$$
$$+ E_{x'} \cdot \left(\cos^2\varphi\sin\varphi\left(r_{13} + 2r_{42}\right) + r_{33}\sin^3\varphi\right) \quad (7)$$

From this expression, we can derive the absolute refractive index $n_{x'x'}$ for light polarized along the $x'$-axis, and the Pockels coefficient $r_{1'1'} = r_{x'x'x'}$ for a modulating field along $x'$ and light polarized along the $x'$ axis

$$n_{x'x'}(\varphi) = \left(\frac{\cos^2\varphi}{n_o^2} + \frac{\sin^2\varphi}{n_e^2}\right)^{-1/2}$$
$$r_{x'x'x'}(\varphi) = -\cos^2\varphi\sin\varphi\left(r_{13} + 2r_{42}\right) - r_{33}\sin^3\varphi \quad (8)$$

### B. MBE-Grown BTO Thin Film

The results derived above hold for monocrystalline BTO. However, in epitaxial thin films on Si, such as grown by MBE, there are three possible orientations of a domain's *c*-axis (along the substrate's *x*-, *y*-, and *z*-axes), each of which can be polarized in anti-parallel directions [45]. As a result, there are six possible ferroelectric polarizations. Most of the film's domains have their *c*-axis in plane and *a*-axis out-of-plane [20], hence they are called *a*-axis films, see Fig. 4(a). The domains' *c*-axes are oriented orthogonally in-plane, a configuration which is due to the epitaxial relation between BTO and its Si substrate [20].

Our interest lays in finding a waveguide orientation, see Fig. 4(b), which maximizes the EO response [46]. The DC bias poles the domains, see Fig. 4(c). Because light travels half of its way in $\varphi = 0°$ domains and half of its way in $\varphi = 90°$ domains, we introduce the average $r_{\mathrm{eff}}(\varphi) =$

$1/2(r_{x'x'}(\varphi) + r_{x'x'}(\varphi + 90°))$, insert Eq. (8) and find that

$$r_{\mathrm{eff}}(\varphi) = \frac{1}{2}\left[r_{33}\left(\cos^3\varphi + \sin^3\varphi\right) + (r_{13} + 2r_{42})\left(\cos\varphi\sin^2\varphi + \sin\varphi\cos^2\varphi\right)\right] \quad (9)$$

This expression is maximal for $\varphi = 45°$ and yields

$$r_{\mathrm{eff}} = \frac{1}{2\sqrt{2}}\left(r_{13} + r_{33} + 2r_{42}\right). \quad (10)$$

For the bulk Pockels coefficients presented in Table I, it is $r_{\mathrm{eff}} = 960\,\mathrm{pm/V}$ for low modulation frequencies and $r_{\mathrm{eff}} = 530\,\mathrm{pm/V}$ for modulation frequencies exceeding 100 MHz.

In-device measurements of the effective Pockels coefficient in epitaxial BTO thin films range between 100 and 360 pm/V [21], [22], [47], [48]. Strain, domain formation, size effects (bulk versus thin film), film morphology, as well as fabrication-related effects can explain the discrepancy to bulk values [49], [50]. Tang *et al.* report an effective Pockels coefficient of approximately 360 pm/V (measured at 1 kHz modulation frequency) for a BTO thin film epitaxially deposited on magnesium oxide (MgO) by metal organic chemical vapor deposition (MOCVD) [47], whereas Girouard *et al.* report 107 pm/V at 30 GHz on the same platform [48]. Xiong *et al.* report 213 pm/V at 1 MHz [21], and Abel *et al.* 300 pm/V at DC, both for films epitaxially deposited on SOI substrate by MBE, the same deposition technique as used in this work. Although these values are smaller than that of BTO bulk material, they are still 3 to 10 times higher than the Pockels coefficient of LNB ($\sim$30 pm/V [17]).

## V. CHARACTERIZATION

The fabricated samples include plasmonic ferroelectric phase shifters and MZMs. The passive characterization reveals propagation losses of the plasmonic waveguides and the performance of the coupling and splitting elements. We further perform active characterization experiments on devices oriented in $\varphi = 45°$ with respect to the crystalline axes and assess the basic EO performance of the PFM. We then investigated the influence of the DC bias voltage, measured the device's EO frequency response and showed temperature resilience of the fabricated devices up to 250 °C. We relate the findings of each experiment to BTO's material properties, to its ferroelectricity and to the origin of its Pockels effect.

### A. Passive Characterization

Fig. 5 shows the passive transmission spectrum of a 10-$\mu$m-long and 70-nm-wide PFM in a spectral window from 1480 up to 1600 nm. The transmission spectrum shows the characteristic ripple of an asymmetric MZM. The 40-nm-wide 3 dB envelope is mainly limited by the wavelength-dependent coupling losses. Extinction ratios exceeding 30 dB and on-chip losses of 23.6 dB (or fiber-to-fiber losses of 35.6 dB because of fiber-to-chip coupling losses of 6 dB per grating coupler) are found for the MZMs.

It is worth mentioning that although our proof-of-principle devices have fairly high losses, the least part of them is fundamental. Experimental results and simulations indicate that a fiber-to-fiber loss of less than 10 dB is in reach: First, the fiber-to-chip coupling losses can be greatly reduced by advanced grating coupler designs. Today, Si photonic grating couplers may offer a coupling loss as low as 1.5 dB [51]. Second, smoother, steeper, and deeper etching of BTO would eliminate sources of excess losses, such as tilt and roughness of the etched sidewalls. Propagation losses can be expected to decrease from ∼1.5 dB/$\mu$m to ∼0.8 dB/$\mu$m. Third, the plasmonic-to-photonic converters as shown in Fig. 1(c) can be designed to minimize loss from 5 dB per coupler to below 1.5 dB. Fourth, the prospective improvement of the voltage-length product makes much shorter devices feasible.

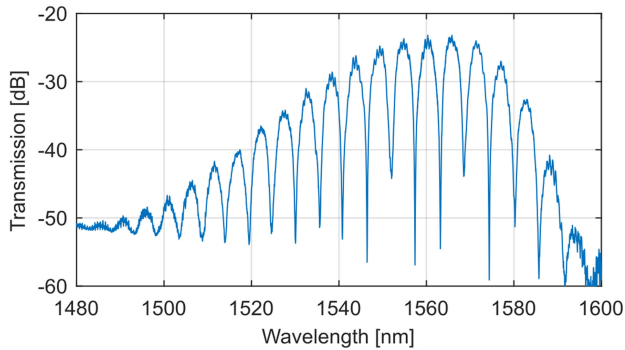

Fig. 5. Transmission spectrum of a fabricated MZM. The active ferroelectric plasmonic section is 10 $\mu$m long and 70 nm wide. 100 $\mu$m path length difference between the two arms results in the observed interference pattern with a free spectral range (FSR) of 5.5 nm. The envelope shape of the spectrum is due to the photonic grating couplers. The spectrum is normalized by the coupling loss of the photonic grating couplers, $2 \cdot 6$ dB = 12 dB.

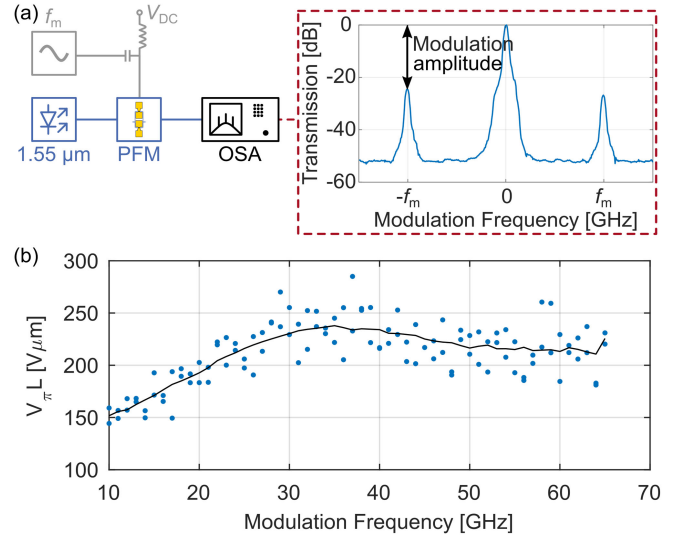

Fig. 6. (a) Measurement setup for the electro-optic device characterization. CW light at 1550 nm, a sinusoidal RF signal and a DC bias are fed to the plasmonic modulator. The optical spectrum is recorded by an OSA. The power ratio between carrier and sideband (indicated by the black arrow) is a measure for the modulator's $V_\pi L$. (b) Extracted voltage-length product over modulation frequency. The solid black line indicates a moving average.

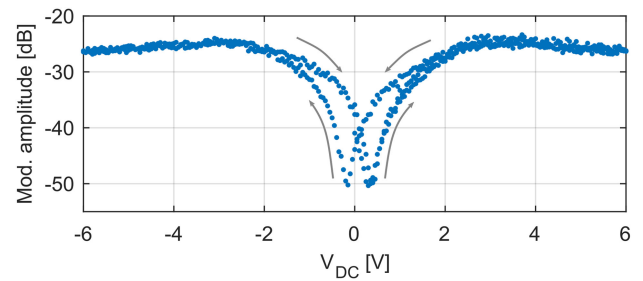

Fig. 7. Hysteresis loop of the sideband-to-peak power ratio. The DC bias voltage has been changed starting from 0 V to +8 V, −8 V, +8 V. The region of interest between -6 V to 6 V is shown. The modulation frequency was 40 GHz, the RF peak-to-peak voltage swing at the device was 2.7 V.

### B. Active Characterization

We have investigated the influence of the DC bias voltage, the RF signal, and the device operation temperature onto the modulation efficiency of the PFM.

The measurement setup for the subsequent experiments is shown in Fig. 6. Light from a continuous wave (CW) laser is fed into a PFM. A bias tee is used to join the modulating RF signal and a DC bias. An optical spectrum analyzer (OSA) detects the optical output signal. The EO modulation causes well-defined optical sidebands, where the power ratio between carrier and sideband (modulation amplitude) is a measure of the phase shifter's modulation efficiency. The modulator's voltage-length product ($V_\pi L$) can be calculated from the modulation amplitude. A detailed explanation and derivation of this analysis method has been reported by Shi and coworkers [52]. We have used phase modulators instead of MZMs, because their basic design rules out unintended influences that more complex devices could introduce. We can conclude, however, that a plasmonic MZM in push-pull mode supports a $V_\pi L$ as small as ∼200 V$\mu$m at a RF signal of 60 GHz, see Fig. 6(b) for the frequency response. From a very recent report on the Pockels coefficients in an MBE-grown BTO film similar to the one used in this work ($r_{42}$ = 932 pm/V, $r_{33}$ = 342 pm/V) [53], we can calculate $r_{x'x'x'}(45°)$ = 750 pm/V. Even assuming a value of only 250 pm/V, and combining it with an im-proved device fabrication (BTO thickness and etch depth both 100 nm), we predict a $V_\pi L$ of below 20 V$\mu$m, at 60 GHz and beyond.

### C. DC Influence and Hysteresis

In this first experiment, we swept the DC bias voltage from −8 to +8 V. The RF signal was tuned at a frequency of 40 GHz and set to an output power of 10 dBm. The modulation amplitude is plotted in Fig. 7. The modulation clearly follows a hysteresis loop, as is expected for ferroelectric materials [54]. This behavior can be explained with the thin film's domain orientation, see Fig. 4. If the applied DC bias field exceeds the domains' coercive field, the directions of the spontaneous polarization are aligned and the modulation efficiency is maximized. If the DC field is reduced and then reversed, this poling is gradually lost, until all spontaneous polarizations are re-aligned in the opposite direction and another maximum is found.

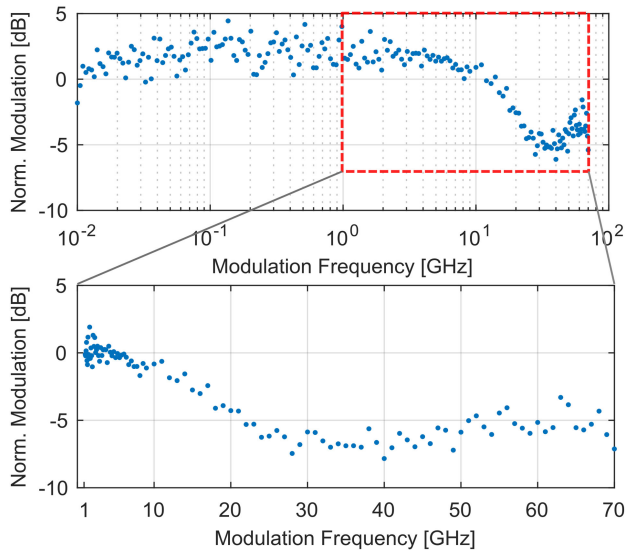

Fig. 8. The measured frequency dependence of the PFM's modulation efficiency follows the frequency dependence of the BTO crystal. Top: Measurement from 10 MHz to 70 GHz on a logarithmic frequency scale. Bottom: The response from 1 GHz to 70 GHz on a linear scale. The influence of strain (distortion of the crystalline lattice) onto the EO response decays in a soft resonance between 10 GHz and 30 GHz. The EO response above this frequency is expected to be flat up to 3 THz [55]. The RF power was 10 dBm at the source, and the DC bias was 2.5 V.

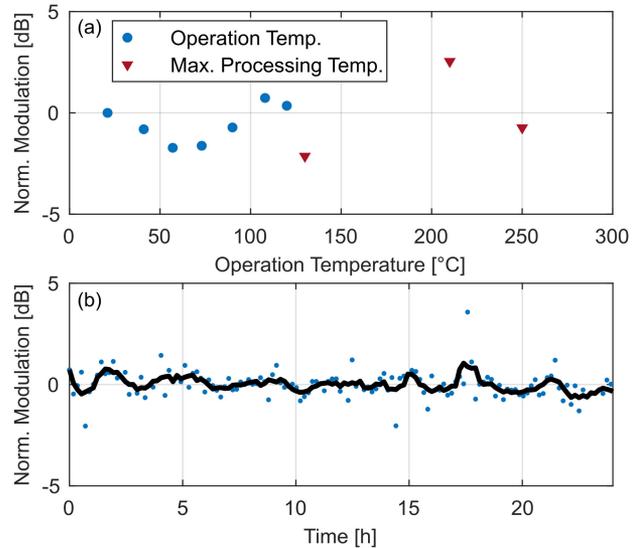

Fig. 9. Investigation of the PFM's temperature stability. The device was operated at 65 GHz. (a) Normalized EO response of a plasmonic phase modulator operated at different temperatures (blue circles), and at room temperature after 5 min heat exposure (red triangles). (b) The normalized EO response of a continuous operation at 90 °C device temperature.

## D. Radio-Frequency Response

We have investigated the frequency response of the PFM from 10 MHz to 70 GHz. In the measurement configuration as described before. We kept the DC bias voltage (2.5 V) and the RF power (10 dBm at the source) constant, while sweeping the frequency of RF signal. The results are shown in Fig. 8, where we have calibrated for the frequency-dependent loss of our RF system. In contrast to other modulator implementations, the plasmonic approach minimizes the influence of the device geometry on the frequency response [32].

Indeed, the measured frequency dependence of the PFM's modulation efficiency follows the frequency dependence of the BTO crystal. We observe a flat frequency response from 10 MHz to ~1 GHz, which drops between 10 and 30 GHz. For higher frequencies up to 70 GHz, the response remains constant.

The observed behavior is expected as reported in the literature [55]–[57]: At low frequencies, BTO's piezo-electric effect causes strain effects (distortions of the lattice) and facilitates a strong EO response. This strong response is reported to decay over two resonances, the first at ~100 MHz, the second – a soft phonon resonance [56] – at ~25 GHz. While we have not observed the first resonance (possibly due to the epitaxial relation between BTO and the Si substrate), the observed second resonance closely follows the predictions for BTO's permittivity dispersion by Laabidi *et al.* [56], which can be related to BTO's EO properties by Fontana *et al.* [57]. A similar behavior for frequencies between 10 GHz and 30 GHz was already reported for BTO thin film modulators on MgO substrate by Girouard *et al.* [48]. It is important to note that BTO's EO response is expected to be constant for frequencies above the acoustic resonances up to 3 THz [55]. This fact is supported by spectroscopic measure-ments of ferroelectric LNB, which exhibits a flat permittivity dispersion from ~10 MHz up to >1 THz [58].

## E. Temperature Stability

The influence of the device temperature on the modulation efficiency has been investigated. To this end, we have subjected the device to temperatures up to 130 °C while measuring the modulation efficiency at 65 GHz. The result is shown in Fig. 9(a). The modulation efficiency does not strongly depend on the temperature. Moreover, the modulation prevails even at temperatures exceeding the Curie temperature of bulk BTO (120 °C [59], 123 °C [60]), at which the bulk material undergoes a phase transition from the tetragonal to a cubic, centrosymmetric phase. Curie temperatures exceeding 650 °C have been reported for epitaxially grown BTO thin films. Strain-induced effects are responsible for this enhancement [61].

In a second part to this experiment, we subjected the device to temperatures up to 250 °C for five minutes each and tested the modulation efficiency after each exposure, see the triangular symbols in Fig. 9(a). No device performance degradation could be found.

To get a first insight about the stability of the device, we subjected it to a continuous operation of 24 hours at 90 °C, a typical operating temperature required for practical applications. The device did not show any sign of degradation, see Fig. 9(b).

## VI. DATA MODULATION EXPERIMENTS

The experimental setup for the data modulation experiments are depicted in Fig. 10. In this experiment, we demonstrated 58 GBd PAM-4 data modulation with a line rate of 116 Gbit/s and 72 GBd NRZ data modulation with a line rate of 72 Gbit/s [23], [62]. Insets show the eye diagrams and signal distribution at different positions of the digital signal processing (DSP) chain.

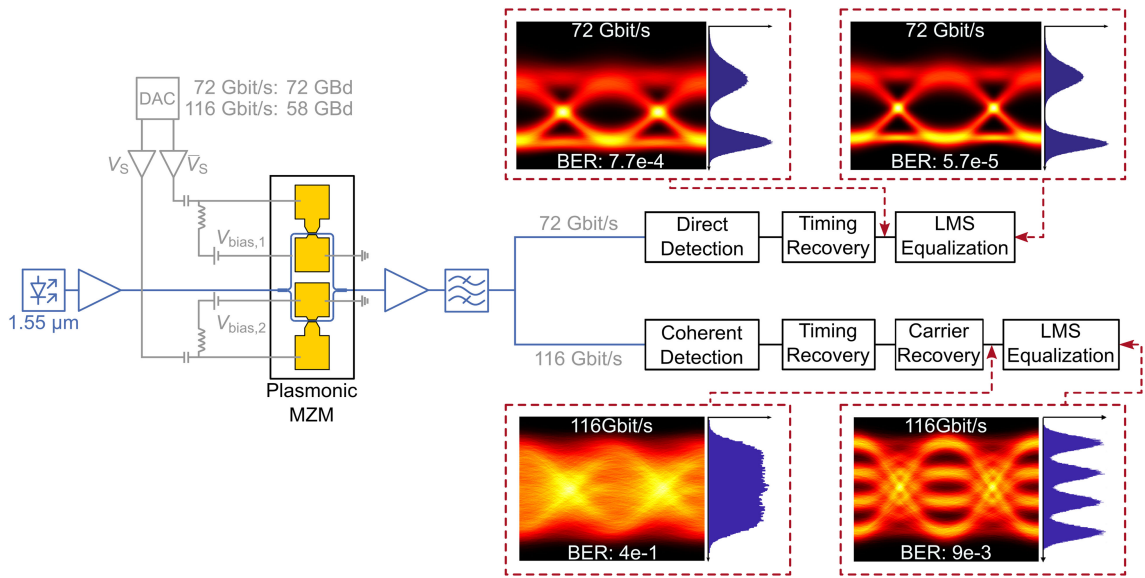

Fig. 10. Schematic of the data modulation experiment. The signal was created offline and sent to a DAC (3 dB BW: <18 GHz). The signal was electrically amplified and fed to the chip. CW light from an ECL was boosted to 18 dBm by an EDFA before it was coupled to the chip. The PFM was driven in push-pull mode. A DC bias was applied on each arm. The modulated light was amplified before it was digitized and processed. The insets show the eye diagrams of both signals before and after equalization.

The signals were generated with a de-Bruijn bit sequence of order 18 and loaded on the memory of a digital-to-analog converter (DAC) with a sampling rate of 72 GSa/s and an electrical 6 dB bandwidth of <29 GHz. The DAC's differential output signals were electrically amplified before they were fed to the chip. The MZM was operated in a dual-drive, push-pull mode. The applied RF voltage peak was $V_p = 2.8$ V, while a DC bias voltage of approximately 3 V was applied over each of the two arms. The applied DC bias ensured that the BTO was fully poled and could additionally be used to fine-tune the operating point of the PFM. Light from an external cavity laser (ECL) at 1.55 μm wavelength was amplified to 18 dBm by an erbium doped fiber amplifier (EDFA) before it was launched onto the chip via a photonic grating coupler. At the output, the optical signal was coupled off-chip to the fiber via another grating coupler and amplified by two cascaded EDFAs.

For data modulation at 72 Gbit/s, the signal was detected by a 70 GHz pin-photodiode and recorded by a real-time oscilloscope (RTO) with a sampling rate of 160 GSa/s and an electrical 3 dB bandwidth of 63 GHz. The digitized signal was processed offline, including timing recovery, adaptive equalization, hard symbol decision and error counting. The adaptive equalizer is a T/2-spaced feed forward equalizer with LMS-based adaptive filter tap updates and 51 filter taps. Finally, we made a hard decision, evaluated $899 \cdot 10^3$ bits, counted the errors and calculated the bit error ratio (BER). A BER of $7.66 \cdot 10^{-4}$ was achieved without any pre- or post-equalization. With post-equalization, a BER of $5.7 \cdot 10^{-5}$ was achieved [23]. Both results are well below a 7% hard-decision FEC threshold of BER = $3.8 \cdot 10^{-3}$ [62].

For data modulation at 116 Gbit/s, the signal was detected by an optical coherent receiver and recorded by the RTO. After timing and carrier recovery, the signal was equalized as described above. Furthermore, we pre-distorted the signal by a non-linear look-up table and compensated for imperfections of the driving electronics. After the hard decision and evaluation of $475 \cdot 10^3$ bits, we determined a BER of $9 \cdot 10^{-3}$ [62]. This is below a 15% soft-decision FEC threshold of $1.9 \cdot 10^{-2}$ [63]. The achieved net data rate exceeds 100 Gbit/s.

## VII. CONCLUSION

We have reported an ultra-compact and ultra-fast PFM. We have characterized the frequency response from 10 MHz to 70 GHz, which remains constant for frequency above ~25 GHz. The EO response in the only 70-nm-wide structured BTO strip remains stable at operation temperatures up to 130 °C. Furthermore, the devices are resilient to temperature exposure up to 250 °C. Data modulation experiments with a MZM at 116 Gbit/s (PAM-4) and 72 Gbit/s (NRZ) have been demonstrated. With recent technology advancement, voltage-length products below 20 Vμm are likely in reach for the presented plasmonic ferroelectric modulator design. In this regime, a device operated with ~1 $V_p$ and with less than 10 dB fiber-to-fiber loss can be realized. The demonstrated performance and robustness of the device, together with the prospective performance improvement, show the technology's great potential for future applications in practical industrial environments.

**Andreas Messner** received the B.Sc. and M.Sc. degrees in electrical engineering and information technology from the Karlsruhe Institute of Technology (KIT), Karlsruhe, Germany, in 2013 and 2015, respectively. He joined ETH Zürich, Zürich, Switzerland, in 2015, and is currently working toward the Ph.D. degree in the group of Prof. Juerg Leuthold.

From 2010 to 2015, he investigated and modeled battery and fuel cell electrode materials as a Student Research Assistant in the Institute of Materials for Electrical and Electronic Engineering, KIT. From December 2013 to May 2014, he was part of the network infrastructure group in the Guiana Space Centre, Kourou, France. His research interests comprise integrated photonics and plasmonics and their combination with nonlinear optical materials.

**Felix Eltes** received the M.Sc. degree in engineering nanoscience from Lund University, Sweden, in 2015. He is a Ph.D. student at IBM Research–Zürich, Switzerland, working toward the Ph.D. degree in materials science at ETH Zürich. His research focus is on 3-D monolithic integration of nonlinear oxides on silicon photonics platforms, aiming to go beyond conventional electro-optic devices for both existing and emerging applications.

**Ping Ma** received the B.E. degree from Tianjin University, Tianjin, China, in 2003, the M.Sc. degree from the Royal Institute of Technology, Stockholm, Sweden, in 2005, and the D.Sc. degree from the Swiss Federal Institute of Technology Zürich (ETH), Zürich, Switzerland, in 2011. His doctoral thesis studied photonic bandgap structures with TM-bandgaps for ultrafast all-optical switches based on intersubband transition in InGaAs/AlAsSb quantum wells.

After a short postdoctoral research stay at ETH Zürich, he was with Oracle Labs in San Diego, CA, USA, where he performed researches on ferroelectric materials for spectral tuning of ring resonators and novel electro-optic modulators for silicon photonic interconnects. He has returned to ETH Zürich and joined in Prof. Juerg Leuthold's institute as a Senior Research Scientist. His current research efforts focus on advanced functional materials and enabled optoelectronic devices for high-speed and low-power-consumption optical communication applications.

**Stefan Abel** studied nanoscale engineering at the University of Würzburg, Germany and received the Ph.D. degree from the University of Grenoble, France. With his background on materials science related to oxide materials and integrated photonics, he codeveloped a hybrid barium titanate/silicon photonic technology. He currently focuses on non-von Neumann computing in the electrical and optical domain, including the implementation of photonic reservoir computing concepts.

**Benedikt Baeuerle** received the B.Sc. and M.Sc. degrees in electrical engineering and information technology from the Karlsruhe Institute of Technology, Karlsruhe, Germany, in 2010 and 2013, respectively, and is currently working toward the Ph.D. degree in the Institute of Electromagnetic Fields, ETH Zürich, Zürich, Switzerland.

From March 2012 to August 2012, he visited the Photonics System Group of the Tyndall National Institute, Cork, Ireland, as a Research Intern. His research interests include digital signal processing, real-time processing, digital coherent transceivers, and optical transmission systems and subsystems.

**Arne Josten** received the M.Sc. degree in electrical engineering and information technology from the Karlsruhe Institute of Technology, Karlsruhe, Germany, in 2013, and is currently working toward the Ph.D. degree at ETH Zürich, advised by Prof. Juerg Leuthold. His research interests evolve around digital signal processing for high-speed optical communication, which he is studying in context of capacity maximization, complexity reduction, and real-time applicability.

**Wolfgang Heni** received the M.Sc. degree in electrical engineering and information technology from the Karlsruhe Institute of Technology, Karlsruhe, Germany, in 2013, and is currently working toward the Ph.D. degree in electrical engineering at ETH Zürich, Zürich, Switzerland, in the group of Prof. Juerg Leuthold.

From April 2012 to September 2012, he visited the Photonics System Group, Tyndall National Institute, Cork, Ireland, as a Research Intern. His research focuses on the in-device optimization and the application of nonlinear optical materials, integrated photonics, plasmonics, and electro-optical devices.

**Daniele Caimi**, biography not available at the time of publication.

**Jean Fompeyrine** received the Engineering degree from the National School of Engineer, Caen, France, and the Ph.D. degree from the University of Bordeaux, France. His expertise relates to functional oxide thin films, used as materials for CMOS and integrated photonics. He has also focused on new methods for the monolithic heterogeneous integration of advanced materials. He is currently focusing on the development of dedicated hardware for neuromorphic computing, specifically novel analog nonvolatile resistance.

**Juerg Leuthold** (F'13) was born in Switzerland in 1966. He received the Ph.D. degree in physics from ETH Zürich for work in the field of integrated optics and all-optical communications.

From 1999 to 2004, he was with Bell Labs, Lucent Technologies, Holmdel, NJ, USA, where he performed device and system research with III/V semiconductor and silicon optical bench materials for applications in high-speed telecommunications. From 2004 to 2013, he was a Full Professor with the Karlsruhe Institute of Technology (KIT), where he headed the Institute of Photonics and Quantum Electronics and the Helmholtz Institute of Microtechnology. Since March 2013, he has been a Full Professor with the Swiss Federal Institute of Technology (ETH).

Dr. Leuthold is a Fellow of the Optical Society of America. When being a Professor with the KIT, he was a member of the Helmholtz Association Think Tank and a member of the Heidelberg Academy of Science. He currently serves as a Board of Director in the Optical Society of America. He has been and is serving the community as general chair and in many technical program committees.